\documentclass[twocolumn,aps,showpacs,preprintnumbers,amsmath,amssymb]{revtex4}
\usepackage{epsfig}
\usepackage{amssymb}

\begin{document}
\def\mydag{^{\vphantom{\dagger}}}
\title{Possibility of Coherent Phenomena like Bloch Oscillations with Single Photons via $W$-States}

\author{Amit Rai and G. S. Agarwal} \affiliation{Department of Physics,
Oklahoma State University, Stillwater, Oklahoma 74078, USA}

\date{\today}

\begin{abstract}
We examine the behavior of single photons at
multiport devices and enquire if coherent effects are possible. In
particular we study how single photons need to be manipulated in
order to study coherent phenomena. We show that single photons need
to be produced in W-states which lead to vanishing mean amplitude
but non\-zero correlations between the inputs at different ports.
Such correlations restore coherent effects with single photons. As a
specific example we demonstrate Bloch oscillations with single
photons and thus provide strict analog of Bloch oscillation of
electrons.
\end{abstract}

\maketitle\section{Introduction} Experiments with single photons
have been at the heart of Quantum optics. About a century ago Taylor
enquired if it is possible to do Young's double slit experiment with
a feeble source of photons \cite{Taylor}. His answer was yes
provided the experiment was done for long enough which was about
$2000$ hours in his case. The interest in single photon interference
experiments has been revived as now we have the possibility of
heralded single photon source
\cite{Alfred,Fasel,Pittman,Bertocchi,Zavatta,Shi}. Variety of other
possibilities with single photon sources have been discussed
\cite{Mandel}. These for example include interaction free
measurements \cite{Elitzur, Kwiat}, Wheeler's delayed choice
measurements \cite{Hellmuth,Jacques} and delayed choice quantum
eraser \cite{Kim}. Application of single photons in quantum imaging
\cite{Sergienko,Teich,Howell,Rubin} have been discussed and single
photons are finding increasing use in quantum information science
\cite{ Walmsley, Brien}. Linear optical elements like beam
splitters; wave\-guides; phase shifters; polarization beam splitter
have been studied to process information with single photons
\cite{Hong, Pathak,Politi,Nagata}. The behavior of such elements is
quite different for single photons and for weak coherent light. The
interpretation of single photon interference experiments is always
intriguing \cite{Feynman, Dirac}. For example the interference with
single photons in a Mach-Zehnder interferometer can be understood in
terms of the \textit{non-factorizability} of the quantum states of
the two mode into which a single mode of light is split by the
action of the beam splitter. Motivated by the above we enquire
generally whether coherent phenomena can be possible with incoherent
single photon sources. We show that one needs to make a $W$-state
\cite{Dur,Eibl} out of a single photon source. This is possible by
the use of a multi\-port beam splitter. The $W$-state has strong
quantum correlations even though it has no coherent component for
the field amplitudes. The quantum correlations in the $W$-state are
responsible for restoring the coherent effects.

The organization of the paper is as follows : In section II, we show
coherent effects can be possible with incoherent single photon
sources. In section III, we demonstrate the possibility of coherent
Bloch oscillations \cite{Bloch,Peschel,Pertsch} using single photon
sources. This fills a gap that has existed as Bloch oscillations
with coherent light fields do not provide strict analog of the Bloch
oscillations for the case of electrons. In the latter case we have a
quantum particle whereas in the case of coherent beam of light we
have a classical source. We summarize our results in section IV and
conclude with future perspectives.

\section{Possibility of coherent Phenomena using Incoherent single Photon $W$-States}

In order to understand the role of coherence and correlations,
consider interference of the two beams of light. Let each beam be
characterized by the annihilation and creation operators
$a_{\alpha}$ and ${a_{\alpha}}^\dagger$; $\alpha$= $1$, $2$. We
write the total field operator after the two beams are made to
interfere as
\begin{eqnarray}
 b_1 &=& ({a_{1}+ e^{i\theta}a_{2}})/{\sqrt{2}}~.
 \label{eq1}
\end{eqnarray}

\begin{figure}[htp]
 \scalebox{0.38}{\includegraphics{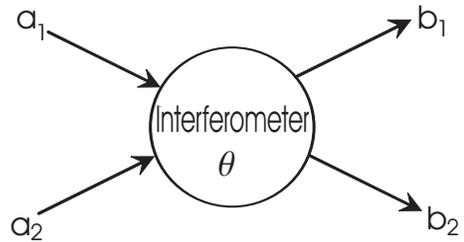}}% Here is how to import EPS art
 \caption{\label{Fig2(a)}Schematic diagram of a two beam interferometer.}
 \end{figure}

\noindent The mean intensity of the output $b_{1}$ is then
\begin{eqnarray}
\langle {b_{1}}^\dagger b_1 \mydag\rangle &=& \frac{\langle
a_{1}^\dagger a_{1}\mydag\rangle + \langle a_{2}^\dagger
a_{2}\mydag\rangle + e^{i\theta} \langle a_{1}^\dagger a_{2}\rangle+
e^{-i\theta} \langle a_{2}^\dagger a_{1}  \rangle}{2}~.\nonumber\\
\end{eqnarray}

\noindent Clearly for interference to occur we need

\begin{eqnarray}
 \langle {a_{1}}^\dagger a_2 \mydag\rangle & \neq & 0~.
\end{eqnarray}

\noindent If the two beams are in coherent states, then

\begin{eqnarray}
 \langle {a_{1}}^\dagger a_2 \mydag\rangle & = &   \langle {a_{1}}^\dagger
\rangle \langle a_2 \mydag\rangle~,
\end{eqnarray}

\noindent and thus interference obviously occurs. If on the other
hand the input state were Fock state $|n_1,n_2\rangle$, then
$\langle {a_{1}}^\dagger a_2 \mydag\rangle = 0$ and no interference
occurs in the mean intensity given by (2). We are required to have
non-zero correlation $(3)$. Thus one can consider an entangled state
of the form

\begin{eqnarray}
 |\psi\rangle &=& \frac{(|1,0\rangle + |0,1\rangle)}{\sqrt{2}}~,
\end{eqnarray}

\noindent then

\begin{eqnarray}
 \langle {a_{1}}^\dagger a_2 \mydag\rangle & = & \frac{1}{2}~.
\end{eqnarray}

\noindent Therefore for the observation of interference at the level
of mean intensity one needs to have non-zero correlation which is
possible with a state like $(5)$ unless one is dealing with coherent
beams of light.

\begin{figure}[htp]
 \scalebox{0.38}{\includegraphics{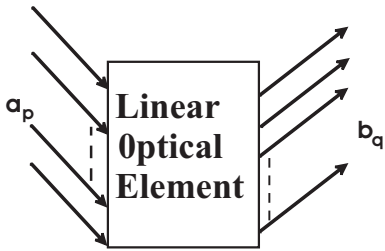}}% Here is how to import EPS art
 \caption{\label{Fig2(a)}The figure shows the linear optical element with input fields $a_{p}$ (for
$p$=1, . . . ,N) and output fields
 $b_{q}$.
 }
 \end{figure}

\noindent Let us consider a more general situation with several
inputs and several outputs for an optical element. We write the
input-output relation as

\begin{eqnarray}
b_{q} & = & \sum_{p=1}^{N}  G_{p, \hspace{0.02 cm} q}\hspace{0.03
cm} a_{p}~,
\end{eqnarray}

\noindent where $G$ depends on the linear optical element and is
equal to $\delta_{p \hspace{0.02 cm} q}$ in the absence of the
optical element. If all the inputs are in coherent states with
amplitudes $\alpha_{p}$, then

\begin{eqnarray}
 \langle b_{q} \mydag\rangle & = & \sum  G_{p, \hspace{0.02 cm}
q}\hspace{0.03 cm} \alpha_{p}~,
\end{eqnarray}

\noindent and

\begin{eqnarray}
 \langle b_{q}^\dagger b_{q} \mydag\rangle & = & | \langle b_{q} \mydag\rangle|^{2} \equiv \Big|\sum_{p}  G_{p, \hspace{0.02 cm}
q}\hspace{0.03 cm} \alpha_{p}\Big|^{2}~.
\end{eqnarray}

\noindent On the other hand if the inputs are incoherent, then we
loose all coherent effects

\begin{eqnarray}
 \langle b_{q}^\dagger b_{q} \mydag\rangle \equiv \sum_{p}  {G}_{p, \hspace{0.01 cm}
q}^* G_{p, \hspace{0.02 cm} q}\hspace{0.03 cm}
 \langle {a_{p}}^\dagger a_{p} \mydag\rangle~.
\end{eqnarray}

\noindent The question is$-$ how can one restore the interference
effects with single photons. In order to see what is needed to
restore the interference effects we use (7) and write the output
intensity in the form

\begin{eqnarray}
 \langle b_{q}^\dagger b_{q} \mydag\rangle \equiv \sum_{r,\hspace{0.02 cm}s}  {G}_{q, \hspace{0.01 cm}
r}^* G_{q, \hspace{0.02 cm} s}\hspace{0.03 cm}
 \langle {a_{r}}^\dagger a_{s} \mydag\rangle~.
\end{eqnarray}

\noindent Therefore for ensuring the interference effects of the
type implied by (9), we need to have

\begin{eqnarray}
 \langle {a_{r}}^\dagger a_s \mydag\rangle & \neq & 0  \hspace{0.2 cm} \forall \hspace{0.1 cm} r,
 s
\end{eqnarray}

\noindent even if $ \langle  a_r \mydag\rangle = 0$. One may be able
to find several states satisfying (12) we have found that if the
input fields are in a $W$ state, then (12) holds. Let us represent
the input single photon state as a $W$ state

\begin{eqnarray}
&& |\psi\rangle \equiv   \sum_{p=1}^{N} c_{p} |1_{p},\{0\}\rangle,\nonumber\\
&& \sum_{p}|c_{p}|^{2}=1~,
\end{eqnarray}

\noindent and where $|1_{p},\{0\}\rangle$ denotes single photon at
the $p^{th}$ input and no photons at the remaining inputs.
Traditionally one takes $c_{p}=1/\sqrt{N}$ but it is not essential.
The $W$-state has the unusual property

\begin{eqnarray}
 \langle a_{p}^\dagger a_{q} \mydag\rangle = c_{p \hspace{0.02 cm}
}\hspace{0.01 cm}^* c_{q} \hspace{0.03 cm}, \langle a_{p} \rangle =
0~;
\end{eqnarray}

\noindent and hence the output (in Eq. (11))  becomes

\begin{eqnarray}
 \langle b_{q}^\dagger b_{q} \mydag\rangle = \Big| \sum_{p} G_{p, \hspace{0.02 cm} q}\hspace{0.03 cm}
c_{p}\Big|^{2} ~.
\end{eqnarray}

\noindent Thus the $W$-state for a single photon behaves like a
coherent input for observations of interference at the level of mean
intensity. We have thus proved that any $W$-state of the form (13)
would exhibit interference phenomena like that exhibited by a
coherent state if the amplitude of the coherent state is substituted
for the quantum amplitudes in the superposition (13).

\begin{figure}[htp]
 \begin{tabular}{cc}
\scalebox{0.3}{\includegraphics{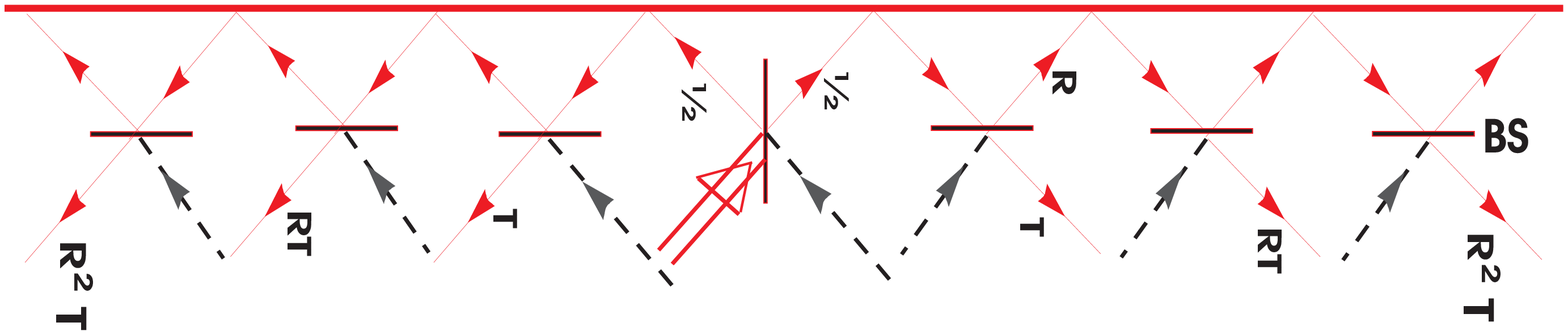}}
 \end{tabular}
 \caption{\label{Fig11}(Color online) Figure shows the scheme for generating the required input
 $W$-state. The thick red line is the mirror with $100    \%$ reflectivity.
 The double arrow
 indicates a heralded single photon from a source like parametric down\-converter. All
black lines show vacuum fields at open ports. The transmissivity of
the beam splitter is $T$. The output intensities at different ports
are given by $---TR^2/2,TR/2,T/2 ,T/2, TR/2, TR^2/2
---$. }
 \end{figure}

 \vspace{0.08 cm}

\noindent We next discuss the arrangement that can produce the
$W$-state (13). In the Fig.\ 3 we show how using a multi\-port
optical splitter \cite{Zukowski, Paternostro} one can produce single
photon $W$-state starting from a heralded single photon source which
has been used by many workers. We show in the Fig.\ 4 the
distribution of $|c_{p}|$ produced by the arrangement of the Fig.\
3. In the next section we present an illustrative example of the use
of the $W$-state in producing coherent effects.

\begin{figure}[htp]
 \begin{tabular}{cc}
\scalebox{0.4}{\includegraphics{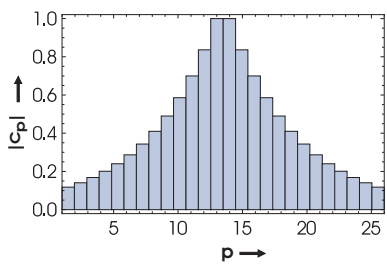}}
 \end{tabular}
 \caption{\label{Fig4} (Color online) Figure shows distribution of $|c_{p}|$ as a function of $p$. }
 \end{figure}

\section{Coherent Bloch Oscillations using Single Photon $W$-states as input to Coupled Wave\-guides}

\noindent The demonstration of Bloch oscillations using optical
elements has attracted considerable attention. It turned out that
such a coherent phenomena which was first discussed in the context
of motion of electrons \cite{Bloch} in a periodic potential and an
electric field can be demonstrated using simple optical structures
and coherent light beams \cite{Pertsch, Peschel}. In view of the
current interest in single photon states it is natural to explore
the possibility of Bloch oscillations with single photons. This
would be strict quantum analog of electronic Bloch oscillations. On
the basis of our discussion in Sec II, we show that Bloch
oscillations with single photons are indeed possible provided we
prepare single photons in a $W$-state.

\noindent We consider an array of $N$ evanescently coupled
single-mode wave\-guides \cite{hagai,Z}, with a linearly varying
refractive index across the array. The mode for the field in the
$p^{th}$ waveguide is described by the annihilation operator
$a_{p}(t)$. These obey Bosonic commutation relations. The
Hamiltonian in terms of the Heisenberg operators can be written in
the form

\begin{eqnarray}
 H = \hbar \sum_{p}   \delta \hspace{0.03cm} p \hspace{0.03cm}a_p^\dagger \hspace{0.05cm} a_p\mydag+ \hbar \sum_{p} J(a_p^\dagger\hspace{0.05cm}
 a_{p+1}\mydag+ a_{p+1}^\dagger\hspace{0.05cm}
 a_{p}\mydag)~,
\end{eqnarray}

\noindent where $J$ is the coupling between the nearest neighbor
wave\-guides and the sum is over nearest neighbors. The refractive
index of $p^{th}$ wave\-guide depends on the index $p$ of the
wave\-guide. Note that in the electron problem the first term in
$(16)$ corresponds to the electric field and the second term to the
periodic potential. The Heisenberg equations of motion are

\begin{eqnarray}
\dot{a}_{p} = -i  \delta \hspace{0.03cm} p \hspace{0.05cm}
a_p\mydag-i J(\hspace{0.05cm}
 a_{p+1}\mydag+ a_{p-1})~,
\end{eqnarray}

\noindent Because of the linearity of the equations (17); the
Heisenberg operators at time $t$ can be expressed in terms of the
operators at time $t=0$

\begin{eqnarray}
a_{p}(t) & = & \sum_{q} G_{p, \hspace{0.02 cm} q}\hspace{0.03 cm}(t)
a_{q}(0)~,
\end{eqnarray}

\noindent where

\begin{eqnarray}
\dot {G}_{p, \hspace{0.02 cm} q}\hspace{0.01 cm}(t) = -i  \delta
\hspace{0.03cm} p \hspace{0.05cm} G_{p, \hspace{0.02 cm} q}\mydag-i
J(\hspace{0.05cm}
 G_{p+1,q}\mydag+ G_{p-1,q})~,\label{eq18}
\end{eqnarray}

\noindent and where the initial condition is

\begin{eqnarray}
G_{p, \hspace{0.01 cm} q}\hspace{0.03 cm}(0) = \delta_{p q}~.
\end{eqnarray}

\noindent It should be borne in mind that the parameter $t$ is
related to the propagation distance by $t=z n/c$ where $n$ is the
refractive index for the mode of the wave\-guide. For large number
of wave\-guides, the Eq.~(\ref{eq18}) can be solved in terms of
Bessel functions. The method of solution is similar to the one in
Ref. \cite{Peschel} and is based on the use of Fourier series
representation. The result can be written as \cite{footnote1}

\begin{widetext}

\begin{eqnarray}
G_{p, \hspace{0.01 cm} q}\hspace{0.02 cm}(t) = \exp \Big[i \alpha q
\tau +\frac{i(p-q)\hspace{0.02cm}(\alpha \tau-\pi)}{2}
 \Big] J_{q-p}[\frac{4}{\alpha} \sin(\frac{\alpha \tau}{2})],\hspace{0.02cm}
\tau\equiv J t, \alpha=\delta/J
~.\nonumber\\
\end{eqnarray}

\end{widetext}

\noindent We can calculate the output for different initial states
of single photon. Consider an arrangement of $2\hspace{0.02cm}N$
wave\-guides. Let single photon be launched in the ${N}^{th}$
wave\-guide. Then the output distribution is given by

\begin{eqnarray}
 I_{p} &=&  \langle a_{p}^\dagger(t) a_{p}(t) \mydag\rangle\nonumber\\
&=&\Big|G_{p, \hspace{0.01 cm} N}(t)\hspace{0.03 cm}
\Big|^{2} \nonumber\\
&=& \Big|J_{p-N}[\frac{4}{\alpha} \sin(\frac{\alpha
\tau}{2})]\Big|^{2}~.
\end{eqnarray}

\begin{figure}[htp]
 \begin{tabular}{cc}
 \scalebox{0.5}{\includegraphics{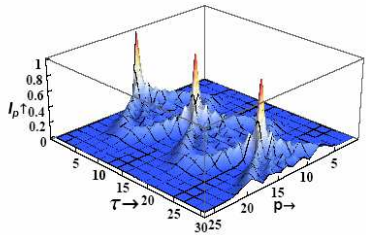}}
 \end{tabular}
 \caption{\label{Fig5} (Color online) Figure shows the variation
of output intensity distribution $I_{p}$ for a single waveguide
excitation as a function of $\tau$ for $p$=1, . . . ,26. The
parameter $\alpha$ is $\alpha=0.5$.}
 \end{figure}

 \begin{figure}[htp]
 \begin{tabular}{cc}
 \scalebox{0.6}{\includegraphics{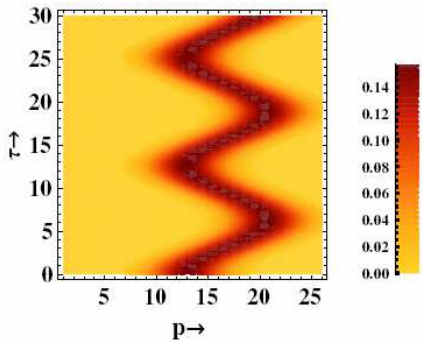}}
 \end{tabular}
 \caption{\label{Fig7}(Color online) Figure shows the Bloch oscillation for a
Gaussian excitation. The number of wave\-guides in the system is
$26$ and the parameter for the Gaussian beam are chosen as
$\sigma=3.6$, and $\bar{p}= 13$. The parameter $\alpha$ is
$\alpha=0.5$.}
 \end{figure}

\noindent This is shown in the Fig.\ 5. The behavior is determined
by the zeroes of the Bessel function.

\vspace{0.08 cm}

 \noindent Next we consider the well known coherent
Bloch oscillation when the input to each wave\-guide is in a
coherent state with amplitude $\alpha_p$. In order to exhibit Bloch
oscillations one needs fairly wide distribution of fields at
different inputs. We assume as in the work of Peschel et al.
\cite{Peschel}, a Gaussian distribution of $\alpha_p$ i.e. we assume
$\alpha_p \sim \exp[-\hspace{0.02cm}{ {(p-\bar{p})}^2}/{2
\hspace{0.05cm} \sigma^2}]$ up\-to a constant. The resulting Bloch
oscillation is shown in the Fig.\ 6.

\noindent Next we show how the quantum correlations in a $W$-state
enable us to obtain coherent Bloch oscillations with single photons.
For this purpose we assume that the input to the wave\-guides is
from the multi\-port device of the Fig.\ 3. The state of the field
at the input would be given by Eq (13) with a distribution of
$c_p$'$s$ given by the Fig.\ 4. The amplitudes $c_p$'$s$ would in
principle have complex phase factors associated with the propagation
distance from the beam splitter BS to the $100$\% mirror and back.
These factors have been set as unity. For single photon in a
$W$-state we get

\begin{eqnarray}
 I_{p} = \Big| \sum_{q} G_{p, \hspace{0.02 cm} q}(t)\hspace{0.03 cm}
c_{q}\Big|^{2} ~,
\end{eqnarray}

\begin{figure}[htp]
 \begin{tabular}{cc}
 \scalebox{0.6}{\includegraphics{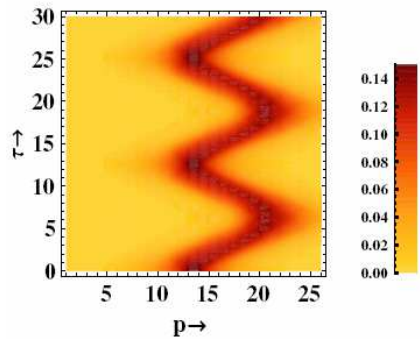}}
 \end{tabular}
 \caption{\label{Fig6}(Color online) Figure shows the Bloch oscillation for a
$W$-state. The parameter $\alpha$ is chosen as $\alpha=0.5$ and the
number of wave\-guides in the system is 26.}
 \end{figure}

\noindent where we use $c_q$'$s$ from the Fig.\ 4. This distribution
of the intensity is shown in the Fig.\ 7. In this case we recover
the coherent Bloch oscillations even though we use incoherent single
photons. This is possible due to the quantum correlations implicit
in the $W$-state of single photons. The similarity between Figs.\ 6
and 7 is striking. This is because of the similarity of the results
(9) and (15). In order to produce the pattern in Fig.\ 6 with single
photons we need an optical device which would produce a distribution
of $c_p$'$s$ (in Eq. (13)) given by a Gaussian distribution.

\noindent An understanding of the Bloch oscillation resulting from
(23) can be obtained by using Fourier space \cite{footnote35}. Let
us write

\begin{eqnarray}
 && c_{q} = \frac{1}{\sqrt{2 \pi}} \int_{-\pi}^{\pi} \tilde{c}(k) \hspace{0.04 cm} e^{i k q} dk\hspace{0.03 cm}~, \nonumber\\
&& \tilde{c}(k) = \frac{1}{\sqrt{2 \pi}} \sum_{q} c_{q} \hspace{0.04
cm} e^{-i k q}
\end{eqnarray}

\begin{widetext}

\noindent Then (23) can be written as

\begin{eqnarray}
 I_{p} = \frac{1}{2 \pi}\Big| \int_{-\pi}^{\pi} \tilde{c}(k-\alpha\hspace{0.03 cm} t)\hspace{0.03 cm}e^{i k p}
 \exp\Big[\frac{2 i}{\alpha}[\sin(k)-\sin(k-\alpha \hspace{0.03 cm}t)]\Big]dk
\Big|^{2} ~,
\end{eqnarray}

\begin{figure}[htp]
 \begin{tabular}{cc}
\scalebox{0.7}{\includegraphics{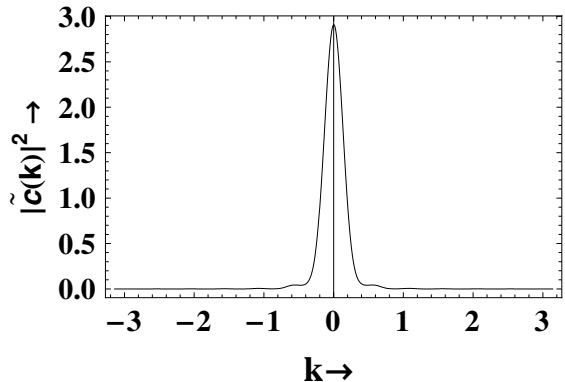}}
 \end{tabular}
 \caption{\label{Fig8} Figure shows the distribution of $|\tilde{c}({k})|^2$ as a function of $k$. }
 \end{figure}

\noindent This in fact was an intermediate step in the derivation of
(21). For the $c_q$ given by the Fig.\ 4, the distribution
$|\tilde{c}({k})|^2$ is shown in the Fig.\ 8. The distribution is
centered at zero and has a width, at half height, $0.31$, which is
quite small in comparison to the range of $k$ values [$-\pi$ to
$\pi$]. Thus an estimate of the behavior of the integral in (25) can
be obtained by expanding the exponential around $\alpha \hspace{0.03
cm}t$ i.e. we set $k \sim (\alpha \hspace{0.03 cm}t+\delta)$ and
approximate

\begin{eqnarray}
  \exp\Big[\frac{2 i}{\alpha}[\sin(k)-\sin(k-\alpha \hspace{0.03 cm}t)]\Big] \approx
 \exp\Big[\frac{2 i}{\alpha} \sin(\alpha \hspace{0.03 cm}t)-\frac{4 i}{\alpha} \hspace{0.03 cm}\delta \sin^2(\frac{\alpha \hspace{0.03 cm}t}{2})
 \Big]
 ~,
\end{eqnarray}

\noindent We have retained terms to lowest order in $\delta$. On
substituting (26) in (25) we get

\begin{eqnarray}
  I_{p} && \cong \frac{1}{2 \pi}\Big| \int \tilde{c}(\delta)\hspace{0.03 cm}e^{i\hspace{0.04 cm}
  \delta \hspace{0.03 cm}(p-\frac{4}{\alpha}\sin^2(\frac{\alpha \hspace{0.03 cm}t}{2}))}
 \exp\Big[i\hspace{0.03 cm} p \hspace{0.03 cm}{\alpha}\hspace{0.03 cm} t +
 \frac{2\hspace{0.03 cm} i}{\alpha} \sin(\alpha\hspace{0.03 cm} t)\Big]\hspace{0.03 cm}d\delta
\Big|^{2} ~, \\ && \cong | c _{\bar{p}}|^{2},\hspace{0.03 cm}
\bar{p} \equiv {\Big( p-\frac{4}{\alpha}\sin^2(\frac{\alpha
t}{2})\Big)}~.
\end{eqnarray}

\noindent The structure shown in the Fig.\ 7 is in conformity with
the approximate formula (28). The revivals in the intensity
distribution are related to the zeros of $\sin^2(\alpha \hspace{0.03
cm}t/2)$.

\end{widetext}

\noindent Finally note that the observation of the single photon
Bloch oscillation would require (a) heralded source of single
photons of the type used in Refs. \cite{Bertocchi, Zavatta, Shi,
Politi} (b) waveguide structures as for example the ones employed in
Refs. \cite{hagai, Z, Politi, iwanow} (c) mirror assembly of the
type discussed by Zukowski et al. \cite{Zukowski}. Since all the
relevant optical elements are currently in use, the observation of
Bloch oscillations with single photons should be possible.

\section{Conclusions}

To conclude, we investigated generally how one can observe the
coherent effects with incoherent single photon sources. For this
purpose one has to convert single photon source into something with
a spatial waveform. We used a multi\-port beam splitter to prepare
single photon $W$-state. Note that recently one has demonstrated
several other interesting possibilities to produce single photon
sources with required waveforms \cite{Yarnall, Kolchin}. Further by
using other types of multiport devices like the ones discussed in
Ref \cite{Zukowski} we can make the magnitudes of all $c_p$ same. As
an application we consider the propagation of light in wave\-guide
array and explored the possibility of observing Bloch oscillations
with single photons. Our results show that the Bloch oscillations
are possible with single photon $W$-state. The quantum correlations
in the $W$-state are responsible for restoring the Bloch
oscillations. There are number of other possibilities using
$W$-state for single photons. For example a phase object in the path
of one of the beams would change the coefficient $c_p$ to $c_p e^{i
\phi_{p}}$ and thus the final interference pattern can be used to
derive information on the object.


\begin{thebibliography}{999}
\bibitem{footnote35} Another possibilty would be to examine the
eigenstates of (16) in the space of single photon excitation. Such
eigenstates can be calculated explicitly. However we have a very
large number of eigenstates and thus an expansion of time dependent
wavefunction in terms of such eigenstates does not turn out to be as
useful as the discussion which leads to Eq. 28.
\end{thebibliography}
\end{document}